\documentclass[%
 reprint,
 amsmath,amssymb,
 aps,
pra,
]{revtex4-2}
\usepackage{xcolor}
\usepackage{graphicx}% Include figure files
\usepackage{floatrow}
\usepackage{subfig}
\usepackage{hyperref}% add hypertext capabilities
\newcommand{\braket}[2]{\langle #1 | #2 \rangle}
\newcommand{\ket}[1]{| #1 \rangle}
\newcommand{\bra}[1]{\langle #1 |}

\begin{document}

%\preprint{APS/123-QED}

\title{Quantum interrogation {of imperfect absorbers} using post-selection}% Force line breaks with \\

\author{Muhammad Abdullah Ijaz}
 % \altaffiliation[Also at ]{}%Lines break automatically or can be forced with \\
\author{Syed Bilal Hyder Shah}
 % \altaffiliation[Also at ]{}%Lines break automatically or can be forced with \\
\author{Muhammad Sabieh Anwar}%
 \email{sabieh@lums.edu.pk}
\affiliation{Department of Physics, Syed Babar Ali School of Science and Engineering, Lahore University of Management Sciences (LUMS), Opposite Sector U, DHA, Lahore 54792, Pakistan.
\\}%

% \collaboration{MUSO Collaboration}%\noaffiliation

%  \homepage{http://www.Second.institution.edu/~Charlie.Author}
% \affiliation{
%  Second institution and/or address\\
%  This line break forced% with \\
% }%
% \affiliation{
%  Third institution, the second for Charlie Author
% \affiliation{%
%  Authors' institution and/or address\\
%  This line break forced with \textbackslash\textbackslash
% }%

% \collaboration{CLEO Collaboration}%\noaffiliation


\begin{abstract}
We propose a scheme for quantum interrogation measurements using constructive interference and post-selection to achieve single-pass high-efficiency detection for imperfect absorbers. We illustrate that our method works for heralded single-photon and weak attenuated sources. We also study the influence of {noise} in our {experimental implementation} and show that post-selection imparts additional robustness to the scheme against noise. We further demonstrate that {fringe visibility} from post-selection interferometry can be used to quantify the transmittance of the imperfect absorber. An interesting link between the interrogation and weak values of nonunitary operators is also highlighted.

% \begin{description}
% Secondary publications and information retrieval purposes.
% You may use the \texttt{description} environment to structure your abstract;
% use the optional argument of the \verb+\item+ command to give the category of each item.
% \end{description}
\end{abstract}

%\keywords{Suggested keywords}%Use show keys class option if keyword
\maketitle

%\tableofcontents

\section{\label{sec:level1} Introduction}

The concept of interaction-free measurement (IFM) can be traced back to Dicke \cite{dicke1981interaction}, who proposed utilizing the non-scattering of the wave function as a measurement. This inspired Elitzur and Vaidman to propose an interferometer scheme using heralded single photons \cite{Elitzur_1993}. In later years, this scheme was experimentally implemented and optimized by including repeated passes through the interferometer to increase detection efficiency \cite{kwiat1995interaction}.

Fortunately, repeated passes also enabled the detection of partially transmitting absorbers. For such absorbers, the object does show coupling with the photon; hence, the general term of quantum interrogation (QI) was introduced \cite{kwiat1999high, geszti1998interaction}, since it is not true IFM. It is important to note that while QI was a consequence of IFM, the formalisms are totally different. The primary difference is the dependence of QI on multiple detections to determine the presence of the object as opposed to the IFM which can achieve this with a single detection albeit still without an interaction.

The first high-efficiency QI scheme was implemented in ultracold atoms and used repeated weak measurements in the polarization degree of freedom of single photons and quantum Zeno stabilization \cite{peise2015interaction}. At about the same time, the IFM scheme was proposed for imaging a class of photosensitive objects \cite{white1998interaction}. Subsequently, the combination of these methods inspired various schemes for interaction-free imaging and detection of a broader range of specimens \cite{hochrainer2022quantum, qian2023quantum, topfer2022quantum} with high efficiencies, which was usually a consequence of repeated passes. Clearly, imaging using undetected photons holds immense promise for quantum sensing, particularly for sensitive biological samples \cite{pualici2022interaction, lemos2014quantum, taylor2013biological}.

However, the efficiency of the imaging methods is bounded by the IFM scheme itself; for example, the efficiency of IFM methods described above relies only on the small, albeit finite, transmittance of the absorber. Therefore, the IFM imaging schemes perform poorly for absorbers with high transmittance. The QI imaging schemes on the other hand, can in principle combat this problem {through repeated passes}. Unfortunately, these schemes also show lower efficiency compared to theoretical estimates as repeated passes through a lossy interferometer accentuate noise \cite{kwiat1999high}.

The scheme proposed in this article considers the absorber's transmittance as well as, importantly, the phase imparted to the photons by the object. Consequently, we can identify the object's presence even if it has high transmittance, with high efficiency even after a \emph{single pass} through the interferometer. Therefore, this method enables the detection of a perfectly transmitting absorber with minimal loss, a distinct advantage compared to other schemes. As such, imaging schemes based on our approach are expected to be more general and will not compromise on efficiency.

The proposed approach also demonstrates robustness against systemic noise sources within the system and is readily adapted to allow for the quantitative extraction of transmittance of the absorber. This possibility is attributable to post-selection embedded into the system and is inherited from the paradigm of weak value via the post selection interferometer (PSI) \cite{fang2018robust, jordan2014technical, rostiom22OptimalSettings}. Traditionally, the weak value technique harnesses pre-selection and post-selection to observe and enhance seemingly minuscule phenomena such as phase shifts \cite{xu2013phase}, magnetic resonance \cite{qu2020sub}, the spin Hall effect \cite{ bai2020impact} and angular rotations \cite{fang2021weak}.

\floatsetup[figure]{style=plain , subcapbesideposition=top}
\begin{figure*}
  \sidesubfloat[]{\includegraphics[scale=0.52]{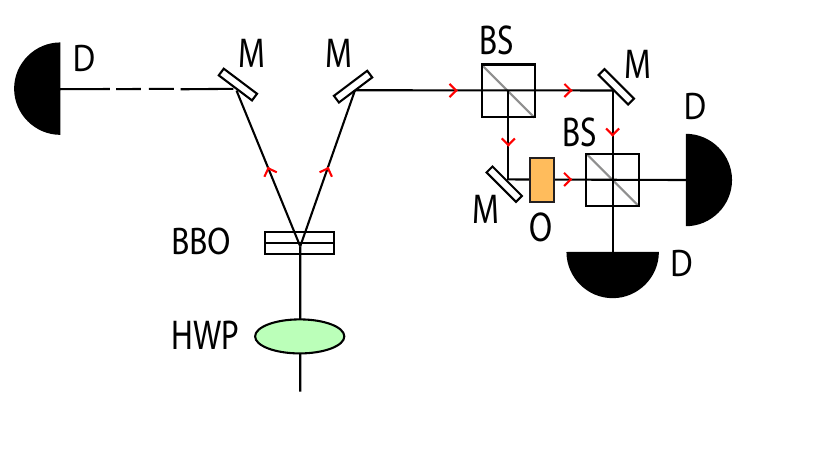}\label{fig:scheme_EV}} \quad
  \sidesubfloat[]{\includegraphics[scale=0.52]{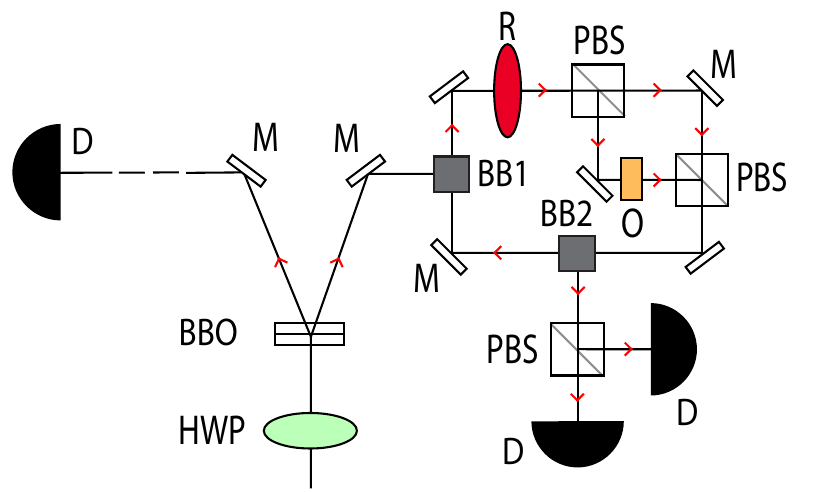}\label{fig:scheme_ZI}}%
  \caption{The schematics given here are variations of the original work that reflect the corresponding theoretical framework, (a) Elitzur-Vaidman's IFM experiment, O (object) is a perfect absorber; (b) interrogation scheme based on the quantum Zeno effect, O (object) has variable transmittance. The components are represented with the following abbreviations: D (avalanche photodiode), BS (beam splitter), PBS (polarizing beam splitter), BBO (beta barium borate crystal), HWP (halfwave plate), M (mirror), BB (black-box) and R (rotator).}\label{fig:schematics}
\end{figure*}

The article is organized as follows. In Sec.~\ref{sec:level2}, we review the evolution of the IFM measurements to high-efficiency quantum interrogation employing the quantum Zeno effect. We introduce our interrogation scheme in Sec.~\ref{sec:level3}, demonstrating {a high theoretical efficiency} for an arbitrarily transmitting object. Then, in Sec.~\ref{sec:level4}, we investigate the influence of technical noise, reflections from optical elements, and offset fluctuations between the interferometer paths on detection probability. In Sec.~\ref{sec:level5}, experimental data is presented, followed by Sec.~\ref{sec:level6} where the setup is used {to measure transmittance of the absorber through post-selection interferometry}. Finally, we conclude in Sec.~\ref{sec:level7}.

\section{\label{sec:level2} Interaction Free Measurement and Zeno-Quantum Interrogation}

Elitzur and Vaidman (EV) proposed the first IFM experiment using the path degree of freedom in a Mach–Zehnder interferometer \cite{Elitzur_1993}. This scheme was realized in $1995$ with single photons \cite{kwiat1995interaction} and later with weak attenuated light and uncharged particles \cite{du1996realization, hafner1997experiment}. The EV setup shown in Fig.~\ref{fig:scheme_EV} utilized the interferometer's path difference such that a single photon emerging from the two output ports of the interferometer underwent constructive and destructive interference. This imposed that the `dark' detector aligned with the destructive port, the `bright' detector aligned with the constructive port, and only the bright detector received detections. This engineered null result utilized the wave nature of light and is only possible in the absence of the object. Now, suppose a perfectly absorbing object (O) is placed in one of the arms of the interferometer, as is shown in Fig.~\ref{fig:scheme_EV}. In this case, the photons traveling in this path project the particle state onto the object or the `dark' and `bright' detectors with equal probability. The efficiency of identifying the object's presence in the path for this EV scheme is defined as $\eta$, a fraction of the `dark' detections over all detections. Since these counts depend on the reflectivity of the BS, the efficiency was upper bound at $1/3$ \cite{Elitzur_1993}.

The original scheme was advanced further to include repeated passes and beam splitters (BSs) with varying reflectivity coefficients to increase the theoretical efficiency for identifying the object's presence \cite{jang1999optical}. For example, by using $N$ {consecutive interferometers (or repeated passes)}, the efficiency took the form,
\begin{align} \label{eq:1}
    \eta = (R)^{N},
\end{align}
where $R = \cos^2(\pi/2N)$ is the reflectivity of each BS and for sufficiently large $N$, $\eta$ was shown to theoretically approach $1$ while the experimental result {was} shown \cite{kwiat1995interaction, white1998interaction} to reach $1/2$.

To overcome this bound, a new scheme was introduced which relied on the polarization degree of freedom and optical quantum Zeno effect \cite{misra1977j, peres1980zeno}, which disrupted the continuous evolution of the photon state when the object coupled with the photon in the system. An equivalent schematic to the original proposed scheme \cite{kwiat1999high} is shown in Fig.~\ref{fig:scheme_ZI}. The circular path composed of the mirrors (M) constitutes the optical loop, which is initially fed with a photon in the horizontal polarization using the black-box (BB1) which may contain i.e. a partially transmitting mirror. In each loop, this photon passes through a rotator (R), which rotates the polarization by $\pi/(2N)$, such that without the presence of the object, the state after $N$ passes is orthogonal to the initial input state. The object is placed inside the arm of an interferometer composed of two polarizing beam splitters (PBS), making it accessible to only the vertical polarization state. Such that in the presence of the object, the vertical polarization component of the photon is coupled to the object. The probability of the object being in the horizontal polarization state after $N$ passes is $\cos^{2N}(\pi/2N)$. Therefore, simply increasing the number of passes through the loop increases the probability of the photon passing through the object-free arm in the interferometer. After the $N^{th}$ loop, the photon is extracted from the system using the blackbox (BB2) containing i.e. a Pockels cell, and its polarization is determined. In the absence of the object, a vertically polarized photon is expected while in the presence of a perfectly absorbing object (O), the vertically polarized component of the photon is absorbed by the object and the wave function of the photon leaving the interferometer in each pass is projected back to the original horizontal state. The consequence of these Zeno measurements is that a photon detected after $N$ passes in the horizontal polarization state signifies the \emph{presence} of the object. The theoretical efficiency of the scheme is $\eta = \cos^{2N}(\pi/2N)$, which approaches $1$ for sufficiently large $N$, but due to the repeated passes through the loop, this experimental efficiency is approximately $3/4$  \cite{kwiat1999high, tsegaye1998efficient}. A key advantage of this setup, however, was that a partially transmitting object could be placed in the interferometer, and measurement of the polarization of the output photon could be used to determine its presence.

\begin{figure*}[]
    \centering
    \includegraphics[width=0.52\textwidth]{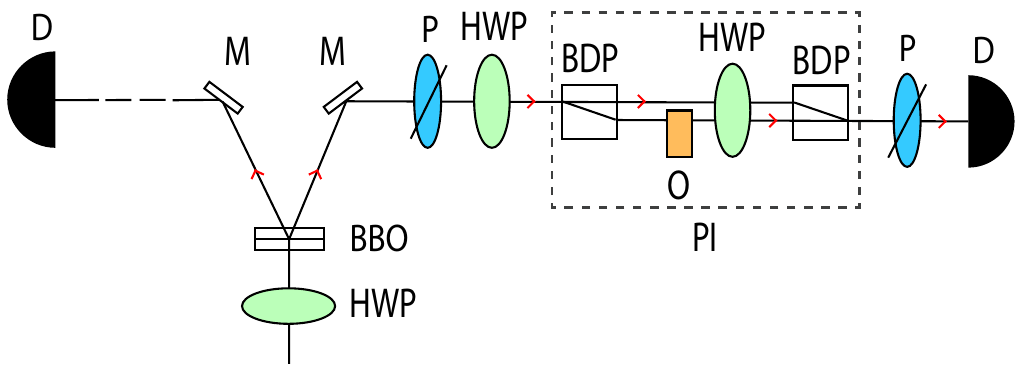}
    \caption{Simple schematic for quantum interrogation using an imperfect absorber O (object). The abbreviations for the optical components are identical to those given in Fig.~\ref{fig:schematics} and include P (linear polarizer) and BDP (beam-displacing prism).}
    \label{fig: schematic}
\end{figure*}

\section{\label{sec:level3} General interrogation Scheme}

The method we present is fundamentally different from most of these earlier schemes. Our strategy removes dependence on the reflectivity of the BS and uses constructive interference instead of destructive interference. The direct consequence of using constructive interference is that our interferometric scheme shifts from IFM to interrogation, such that a single measurement is not sufficient to detect the presence of the absorber, but \emph{can} detect an imperfect absorber with high efficiency. The experimental setup for the proposed scheme is shown in Fig.~\ref{fig: schematic}. It exploits the polarization degree of freedom in the interferometer referenced henceforth as ``polarization interferometer'' (PI), comprising a half-wave plate (HWP) sandwiched between two beam-displacing prisms (BDPs). The intervening HWP equalizes the path length between the polarization states of the single photons \cite{waseem2020, beck2012quantum}. The path variability is achieved by tilting the second BDP about the vertical axis, resulting in the two polarizations adopting variable lengths. The size of the interference fringes quantifies the extent of this interference, which is referred to as the `visibility' and is defined as
\begin{align} \label{eq:v}
    V = \frac{D_{max} - D_{min}}{D_{max} + D_{min}},
\end{align}
where $D_{max}$ and $D_{min}$ are maximum and minimum counts, respectively. 

Furthermore, visibility shows a well-documented complementary relationship with the `which-path' information available to the observer \cite{ma2013quantum, herzog1995complementarity, zou1991induced, rostiom22OptimalSettings}. When the object is placed in one arm of the PI, the single photons in the object-bearing path interact with the object; hence, the object's presence is coupled to the which-path information and, by extension, to the visibility.

\floatsetup[figure]{style=plain} %,subcapbesideposition=left}
\begin{figure*}
  \sidesubfloat[]{\includegraphics[scale=0.48]{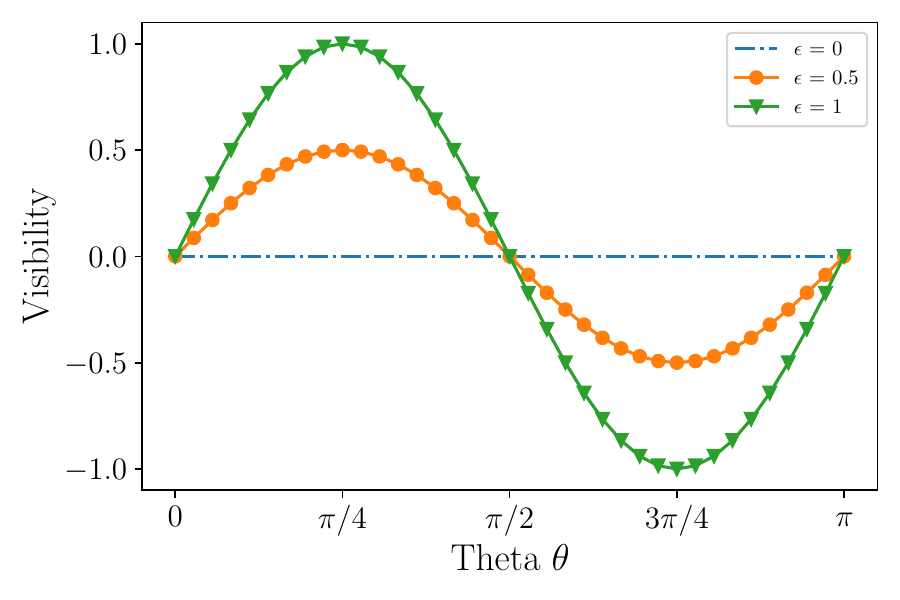}\label{fig:vis_pt}} \quad
  \sidesubfloat[]{\includegraphics[scale=0.48]{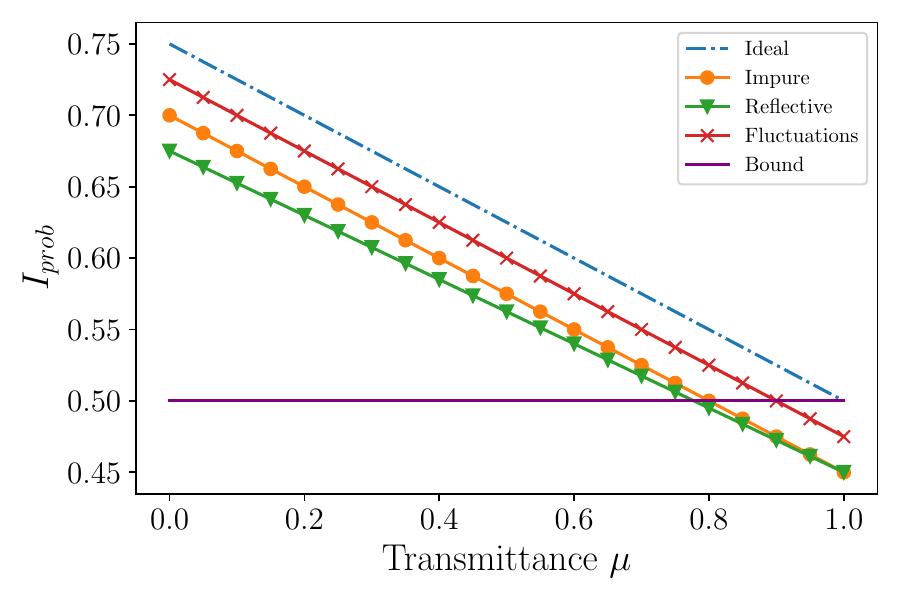}\label{fig:Iprob_noise}}%
  \caption{Theoretical plots for (a) The dependence of visibility on purity and the orientation $\theta$ of the post-selection linear polarizer. The dotted dashed blue curve is for a perfectly mixed pure state, the dotted orange curve is for $\varepsilon = 0.5$, and the green triangle curve is for a pure input state. (b) The probability of identifying the absorber against its transmittance, given an ideal lossless system represented by the dotted dashed blue curve. The graph also shows results for noise from impure input state with $\varepsilon = 0.9$ by the orange dotted curve, from the reflectivity of optical elements $\lambda = 0.1$ with the green triangle curve, phase fluctuations ${\Delta \phi}^{2}= 0.1$ with the red crosses and finally the lower bound of the scheme's efficiency with a solid purple line.}\label{fig:theoretical}
\end{figure*}

In the setup, we employ spontaneous parametric down-conversion (SPDC) using a beta barium borate (BBO) crystal as the single-photon source and a continuously graded neutral density (ND) filter attached to a motorized translation stage as the object. The filter poses variable transmitivity and acts as a semitransparent or partially absorbing) object. The schematics for the previous schemes shown in Fig.~\ref{fig:schematics} share the photon sources with the present scheme for a fair juxtaposition. One output (signal) of the photon from the BBO behaves as a single-photon source when the other photon output (idler) acts as the herald. Alternatively, a highly attenuated light source is achieved if this heralding is not taken into account. If we ignore the heralding counts, the stream of photons behaves as a coherent light source $\ket{\alpha}$ albeit with a very small photon count number $|\alpha|=\sqrt{\overline{n}}$ where $\overline{n}$ is the photon count rate~\cite{loudon2000}. In the subsequent article, we identify this scenario by the phrase ``coherent beam''.

The polarization encodes the quantum state of the single photon, and a straightforward mathematical formulation based on the density matrix describes the predicted experimental outcomes. The initial density state after the pre selection {using the polarizer (P)} oriented along the horizontal (see Fig.~\ref{fig: schematic}) can be represented as \[\hat{\rho_{o}} = \varepsilon \ket{H}\bra{H} + (1-\varepsilon) \frac{\hat{I}}{2},\] where $\varepsilon$ is the state purity. In this manuscript, $\ket{H}$ and $\ket{V}$ represent the horizontal and vertical polarization quantum states. The next component to act on this quantum state is the half-wave plate (HWP) with its optical axis at $\pi/8$, which converts $\hat{\rho}_{o}$ to \[\hat{\rho}_{1} = \varepsilon \Bigg(\frac{\ket{H} + \ket{V}}{\sqrt{2}}\Bigg)\Bigg(\frac{\bra{H} + \bra{V}}{\sqrt{2}}\Bigg) + (1-\varepsilon) \frac{\hat{I}}{2},\] possessing an equal amplitude to take either of the two arms in the PI. In the computational basis $\hat{\rho}_{1}$ can be written as,
\begin{align}\label{eq:3}
    \hat{\rho}_{1}  = \frac{1}{2}
    \begin{pmatrix}
        1 & \varepsilon \\
        \varepsilon & 1
    \end{pmatrix}.
\end{align}
Upon entering the PI, the BDP creates a superposition whilst adding a relative phase. The {action of} the first BDP whose operation is described by \[\hat{O}_{B}(\phi_{1})\ = \ket{H}\bra{H} + e^{i \phi_{1}}\ket{V}\bra{V} \] yields the state,
\begin{align} \label{eq:4}
    \hat{\rho}_{2} = \frac{1}{2}
    \begin{pmatrix}
        1 & \varepsilon e^{-i \phi_{1}} \\
        \varepsilon e^{i \phi_{1}} & 1
    \end{pmatrix}.
\end{align}

\noindent The object of interest is placed in one of the two arms between the BDPs and is represented with a nonunitary $\hat{A}$,
\begin{align} \label{eq:A}
    \hat{A} =
    \begin{pmatrix}
        1 & 0 \\
        0 & e^{i\delta} \sqrt{\mu}
    \end{pmatrix},
\end{align}
where $\mu$ and $\delta$ are the absorber's transmittance and relative phase respectively \cite{azuma2006interaction, geometricphases2000}, given the condition $0\leq \mu \leq1$. Hence, $\hat{\rho_{3}}$, takes the form,
\begin{align} \label{eq:5}
    \hat{\rho}_{3}  =  \frac{1}{2}
    \begin{pmatrix}
        1 & e^{-i (\phi_{1}+\delta)} \varepsilon\sqrt{\mu} \\
        e^{i (\phi_{1}+\delta)} \varepsilon\sqrt{\mu} & \mu
    \end{pmatrix}.
\end{align}
The state $\hat{\rho}_3$ is achieved by replacing the elements in $\hat{\rho}_2$ through the prescription $\ket{V}\rightarrow e^{i\delta}\sqrt{\mu}\ket{V}$ and correspondingly for the adjoint.

The object is followed by the HWP plate with its optical axis at $\pi/4$, {swapping} the polarization states in the PI arms such that at the second  BDP, the paths align, and finally we get the post-interferometer density matrix,
\begin{align}\label{eq:6}
    \hat{\rho}_{4}  =   \frac{1}{2}
    \begin{pmatrix}
        \mu &  e^{i (\phi+ \delta)} \varepsilon \sqrt{\mu}   \\
        e^{-i (\phi+ \delta)} \varepsilon \sqrt{\mu} & 1
    \end{pmatrix}.
\end{align}

\noindent The relative phase $\phi$ is the collection of the phases added by the two BDPs {($\phi_1$ and $\phi_2$)} such that $ \phi = \phi_{1} + \phi_{2}$. Although $\hat{\rho}_{4}$ represents the states after the photon recovers from its superposition state, we need to add another linear polarizer P to fetch us the interference and will function as the post-selector. The operator for P, from Jones calculus, takes the form,
\begin{align} \label{eq:6.1}
    \hat{P} &=
    \begin{pmatrix}
        \cos^2{\theta} &  \cos{\theta}\sin{\theta}  \\
        \cos{\theta}\sin{\theta}  & \sin^2{\theta}
    \end{pmatrix}.
\end{align}

\noindent Without the presence of the absorber ($\mu = 1, \delta = 0$), the probability of the photon passing through the post-selection P for an ideal system ($\varepsilon=1$) is,
\begin{align} \label{eq:6.2}
    Tr[\hat{\rho}_{4} \hat{P}] = \frac{1}{2} \bigg(1 + \sin(2 \theta) \cos(\phi)\bigg).
\end{align}

\noindent Incorporating this into Eq.~\eqref{eq:v}, we find that visibility without the absorber is maximized, as expected when $\theta = \pi/4$. This scenario is shown in Fig.~\ref{fig:vis_pt}. At this orientation, the post-selection polarizer has the operator \[\hat{P}^{'} = \Bigg(\frac{\ket{H} + \ket{V}}{\sqrt{2}}\Bigg)\Bigg(\frac{\bra{H} + \bra{V}}{\sqrt{2}}\Bigg).\]

\noindent On the other hand, in the presence of the absorber the eventual probability of detecting the state after passing through the post-selection polarizer becomes,
\begin{align} 
    D(\mu, \varepsilon, \phi+ \delta) &= Tr[\hat{\rho_{4}} \hat{P}^{'}] \nonumber\\
    &= \frac{1}{4} \bigg(1 + \mu + 2 \varepsilon \sqrt{\mu} \cos(\phi+ \delta)\bigg), \label{eq:D-prob}
\end{align}
which depends on the transmittance $\mu$, phase $\delta$ of the object, purity of the input state $\varepsilon$, and the collective relative phase $\phi$ between the paths inside the PI.

As mentioned above, the proposed scheme relies on constructive interference instead of the destructive interference modality used in EV's work, as this enables higher detection probability for interrogation. To achieve constructive interference, we adjust the {tilt} of the BDPs, hence the PI path lengths, such that the detector outputs $D_{max}$ without the presence of the absorber. Assuming a pure $\hat{\rho}_{o}(\varepsilon=1$), the detections $D_{max}$ should be unity for this orientation and the relative phase $\phi$ should be very small. At this point, subsequent to determining $D_{max}$, we introduce an object in one path of the interferometer, causing the phase between the two paths to become $\phi + \delta$. Since $\delta$ is non-zero, the visibility fringes are shifted and the collective phase between the arms of the PI becomes much larger than the coherence length of the photons \cite{jelezko2003coherence} and we obtain $\cos(\phi+ \delta)\approx 0$. Hence, the probability of detection with an object in the path simplifies to,
\begin{align} \label{eq:8} % make this 9
    D(\mu) &= Tr[\hat{\rho_{4}} \hat{P}] \\
    &= \frac{1}{4} (1 + \mu ). \nonumber
\end{align}
which is seen to be independent of the purity and the object's phase. We propose using the \emph{decrease} in detection probability when an imperfect object is placed in the path of an interferometer that is set to output constructive interference without the object. Therefore, we \emph{define} the probability of successfully detecting the presence of an imperfect absorber as,
\begin{align} \label{eq:9}
    I_{prob} &= D_{max}(\mu = 1, \varepsilon) - D(\mu) \nonumber \\
    & = \frac{1}{2} (1 + \varepsilon)  -  \frac{1}{4} (1 + \mu )  \nonumber \\
    & = \frac{1}{4} ( 1 + 2 \varepsilon - \mu ).
\end{align}

\noindent For a perfect absorber with $\mu = 0$, we have the theoretical detection probability of $3/4$, while for a perfect transmitter $\mu = 1$, the efficiency is $1/2$, given a pure input state. It is also important to note that the metric for efficiency used for this scheme differs from that used for evaluating the EV and later setups \cite{Elitzur_1993, kwiat1995interaction}. Regardless, prior schemes either require a perfect absorber (IFM) {making them unfit} for transmitting objects or incur high losses due to repeated passes (QI) for detecting the presence of a near-perfect transmitter $\mu \approx 1$. The advanced ability of the proposed scheme to detect the presence of the highly transmitting object is, therefore, clear.

\section{\label{sec:level4} Imperfection Analysis}

In this section, we evaluate the robustness of this scheme against two kinds of noise. First, we investigate the impact of reflections from imperfect optical elements, and second, the fluctuations to the relative phase difference $\phi$ in the paths of the PI \cite{fang2018robust}.

\floatsetup[figure]{style=plain} %,subcapbesideposition=left}
\begin{figure*}
  \sidesubfloat[]{\includegraphics[scale = 0.48]{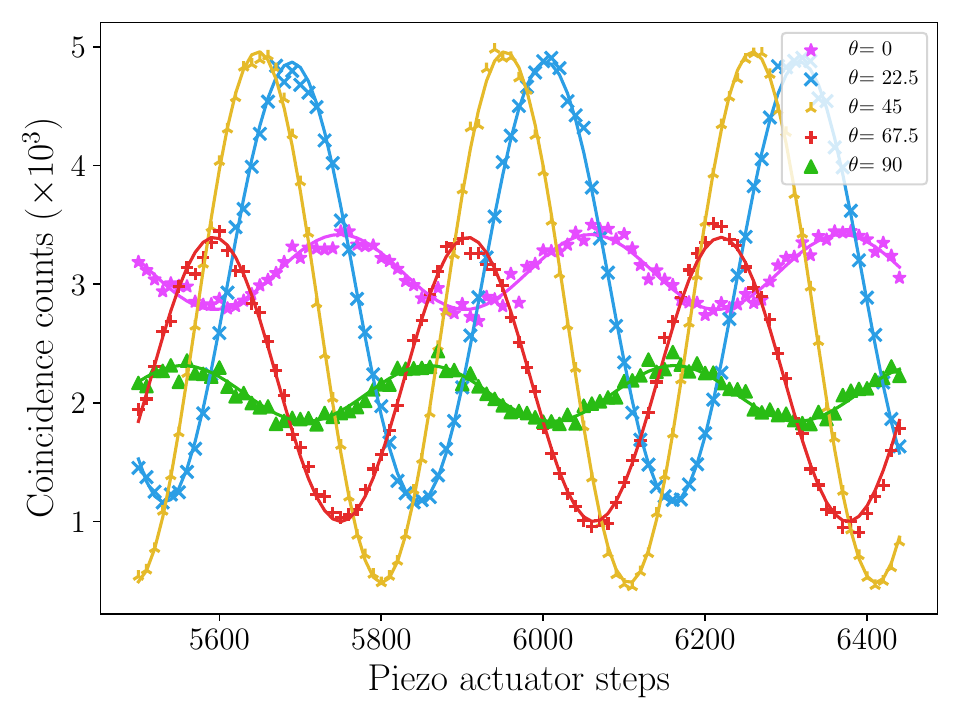}\label{fig:fringes_coin}} \quad
  \sidesubfloat[]{\includegraphics[scale = 0.48]{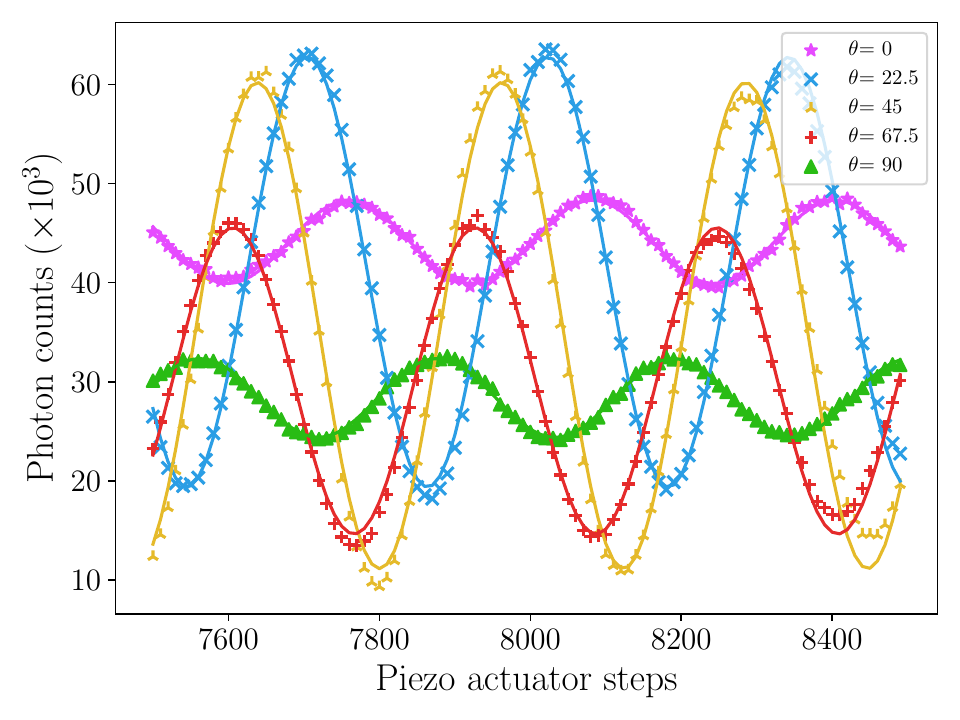}\label{fig:fringes_weak}}%
  \caption{(a) Visibility fringes from the single photons in the absence of an object; (b) Visibility fringes from the coherent beam of light without the presence of the object against the phase induced by the piezo-electric motor. For the diagrams above, the visibility fringes are shown as purple stars for post-selection LP set to $\theta = 0$, blue crosses for $\theta = \pi/8$, yellow tri-up for $\theta = \pi/4$, red pluses for $\theta = 3\pi/8$ and finally green triangles for $\theta = \pi/2$. For visual clarity, results with maximum fringes are shown with a shift added manually. }\label{fig:fringes}
\end{figure*}

The derivation in the previous section assumes that the reflectivity of all optical elements (aside from the absorber) is zero. In reality, the actual operation of the $j^{th}$ element can be represented as $\lambda_{j} e^{i \delta_{j}}$, where $\lambda_{j}$ is the reflectivity and $\delta_{j}$ is the erroneous relative phase introduced. To quantify the influence of reflections, we introduce another quantum state corresponding to reflected photons. This approach is similar to treating IFM through scattering theory, although the additional state does not correspond to absorbed photons \cite{geszti1998interaction, azuma2006interaction}. For this purpose we introduce a new density matrix composed of quantum states propagating forward and backward inside the setup, the latter caused by reflections. Using this description and the assumption that $\lambda_{j} \ll 1$, the state $ \hat{\rho_{i}}$ that has passed through the $j^{th}$ components becomes,
\begin{align} \label{eq:12}
    \hat{\rho_{i}}^{'} &=
    \begin{pmatrix}
        (1 - \sum_{j}\lambda_{j})\hat{\rho_{i}} & 0  \\
        0  & \sum_{j} \lambda_{j}
    \end{pmatrix}.
\end{align}
For simplification, we ignore $ e^{i \delta_{j}}$ as it acts as a phase globally. We can combine the probability of reflected photons from each component to $\lambda$, such that $\lambda =  \prod\limits_{j} \lambda_{j}$. The detection in Eq.~\ref{eq:D-prob} then takes the form,
\begin{align} \label{eq:13}
    D(\mu, &\varepsilon, \phi) =  \frac{1}{4} (1 -\lambda )(1 + \mu + 2 \varepsilon \sqrt{\mu} \cos\phi) ,
\end{align}
and the probability of successfully identifying the presence of the absorber changes from Eq.~\eqref{eq:9} to,
\begin{align} \label{eq:13.5}
    I_{prob}(\lambda) = \frac{1}{4} \bigg((1 - \lambda)(1 + 2 \varepsilon - \mu)\bigg)
\end{align}
which shows a linear dependence on the collective reflectivity of the system and merely a scaling when compared with the idealized case, Eq.~\eqref{eq:9}. This dependence is demonstrated in the simulation results shown in Fig.~\ref{fig:Iprob_noise}, along with the ideal $I_{prob}$ for comparison. Observe that with this kind of noise, $I_{prob}$ can be degraded to less than the theoretical minimum efficiency of $1/2$.

The other noise source vis-à-vis fluctuations in the relative phase $\phi$ is induced by environmental factors such as vibrations, air movements, and deformation of optical mounts. As before, we assume that these fluctuations are small, $\Delta \phi \ll 1$, so we can use the Taylor expansion to second order on the cosine term in Eq.~\eqref{eq:D-prob} about the maximum point to obtain,
\begin{align} \label{eq:14}
    D(\mu,& \varepsilon)_{max} =  \frac{1}{4} \bigg(1 + \mu + 2 \varepsilon \sqrt{\mu} (1 - \frac{{\Delta \phi}^{2}}{2}) \bigg).
\end{align}

\noindent The presence of these fluctuations has a quadratic impact on our scheme, as the probability of successfully identifying the presence of the absorber becomes,
\begin{align} \label{eq:15}
    I_{prob}(\Delta) = \frac{1}{4}( 1 + 2 \varepsilon - \mu - {\Delta \phi}^{2} ).
\end{align}
The simulation results in Fig.~\ref{fig:Iprob_noise} show the impact of these fluctuations on $I_{prob}$.

As shown, the noise has a limited but still observable impact on the scheme's efficiency due to the structural similarity of our experimental arrangement with post-selected interferometry \cite{fang2018robust, jordan2014technical, fang2016ultra}. In fact, the pre-selection and post-selection linear polarizer assumed to be noiseless, enclosing the interferometer filter out the noisy input and output states, making the scheme additionally robust against noise.

\section{\label{sec:level5} Experimental Results for interrogation}

The proposed scheme can function using both single-photon and {coherent} light sources. The construction of the single photon source uses the Type I down-conversion through the BBO at 405 nm, and this same infrastructure provides the coherent beam if the heralding path, which contains only the detector, is blocked or its outcomes are discarded~\cite{waseem2020}. This coherent beam can be extremely sparse due to low photon counts but still shows Poissonian statistics and hence possesses a distinctive classical nature. Using these quantum and classical sources, we assess the practically achievable capabilities of our scheme.

\floatsetup[figure]{style=plain} %,subcapbesideposition=left}
\begin{figure*}[]
  \sidesubfloat[]{\includegraphics[scale=0.48]{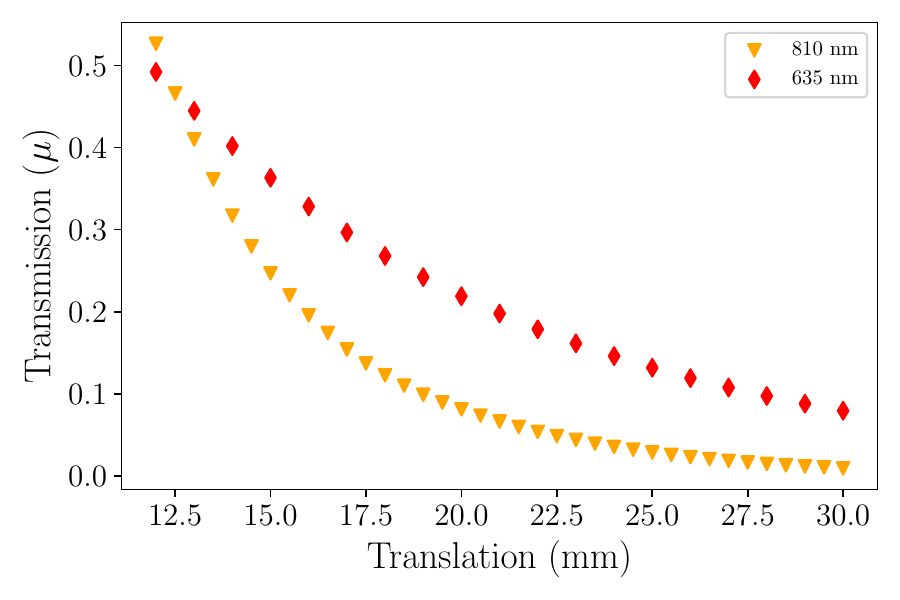}\label{fig:trans_fit}} \quad
  \sidesubfloat[]{\includegraphics[scale=0.48]{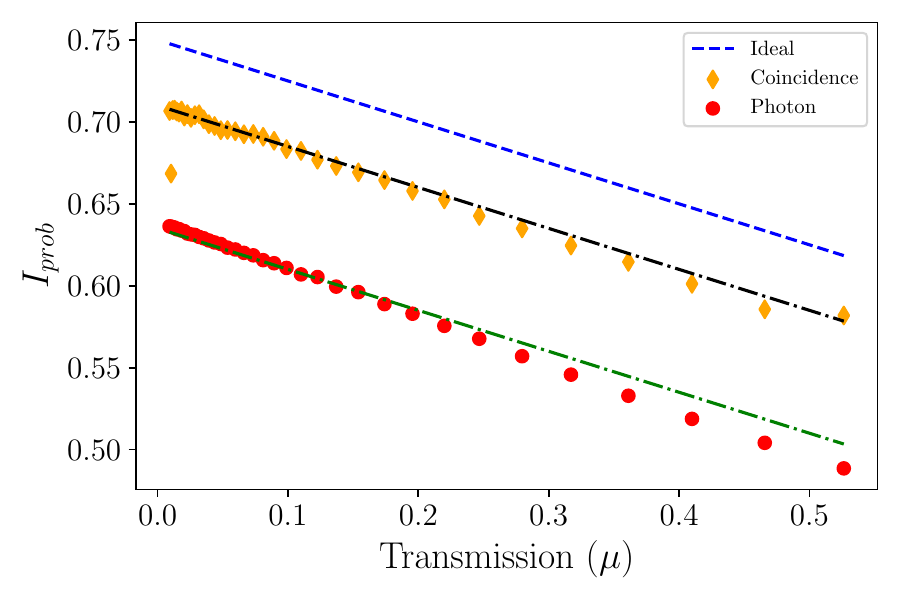}\label{fig:Iprob_trans}}%
  \caption{(a) The relation of transmission of the ND filter with the motion of the actuator stage for $810$ nm represented with downward orange triangles and $635$ nm with red diamonds; and (b) the probability of detecting the presence of the absorber with the transmittance $\mu$ using heralded single photons and coherent light, represented with orange diamonds and red dots respectively. These results are curve fit with a dotted-dashed line in black for a single photon and green for coherent light, with appropriate values of $\varepsilon$, and the ideal result with $\varepsilon= 1$ is shown with the blue dashed line.}\label{fig:experimental}
\end{figure*}

For the setup, we orient the pre-selection polarizer to obtain $\hat{P} = \ket{H}\bra{H}$ ensuring that $\hat{\rho_{0}}$ is approximately a pure state. With this configuration and the absence of the absorber, we achieve the fringes shown in Fig.~\ref{fig:fringes_coin}. The visibility and its respective standard error corresponding to these fringes originating from the different orientations of the post-selection P are shown in Table.~\ref {tab:visibility}. The theoretical and experimental results show that for maximum efficiency, post-selection P should be oriented at $\pi/4$. Therefore, the post- and pre-selection states in our setup are not orthogonal, as is predicted in conventional weak value amplification. 

\begin{table}[b]%The best place to locate the table environment is directly after its first reference in text
\caption{\label{tab:visibility}%
The visibility for different orientations of the post-selection P, given heralded single photons and coherent light. These results are produced in the absence of the object. In each value, the uncertainty in the last digit is shown in brackets.
}
\begin{ruledtabular}
\begin{tabular}{ccc}
% &\multicolumn{2}{c}{Visibility} \\
Orientation ($\theta$) & Single Photon & Coherent Beam \\
\colrule
0      & 0.2232(5) & 0.1121(1) \\
$\pi/8$  & 0.6550(6) & 0.5435(2) \\
$\pi/4$  & 0.8827(8) & 0.7713(2) \\
$3\pi/8$ & 0.6460(7) & 0.5420(2) \\
$\pi/2$  & 0.2817(7) & 0.1586(2) \\
\end{tabular}
\end{ruledtabular}
\end{table}

The deviation of visibility from unity for $\theta = \pi/4$ results from accumulating errors such as reflectivity inherent to the equipment used in the setup and the impurity of the input state. The visibility for the heralded single photons is larger than that for the coherent light source under the same conditions and the same setup shown in Fig.~\ref{fig:fringes_weak}. The coherence lengths, which reflect the coherence time of the excited state for the emitting quantum system and are derived from Heisenberg's uncertainty principle \cite{jelezko2003coherence}, are also larger for the single photon source compared to the coherent beam displaying classical statistics. The interference phenomenon is shared with the coherent light source but has lower visibility, an expected consequence of the higher coherence of single photons compared to classical sources of light \cite{biswas2017interferometric}.

The setup employs a continuously variable neutral density (ND) filter connected to a motorized translational stage as the imperfect object. The actuator stage was experimentally measured to allow for variable transmission $\mu$ between $0$ to $0.526$. The relation of $\mu$ and the filter's position is shown in Fig.~\ref{fig:trans_fit} for $635$ nm as red diamonds and $810$ nm as orange triangles. As shown, the absorptive filter provides various transmission values, each of which is used to calculate the $I_{prob}$, results for which are shown in Fig.~\ref{fig:Iprob_trans}. The orange diamonds represent the experimental data points collected for a heralded single-photon source, while the red dotted curve is for a coherent beam. The absorber detection probability for both sources is lower than theoretical expectations under ideal conditions, as shown in Fig.~\ref{fig:Iprob_noise} with the blue dashed line.

This discrepancy is primarily attributed to the purity of the input state, which, as shown in Eq.~\eqref{eq:9}, has a linear influence on the probability efficiency and is required to be unity for maximum theoretical efficiency. Through a minimization process, we have minimized root mean square error with $\varepsilon$ as the running parameter to fit Eq.~\eqref{eq:9} to the experimental data for heralded single-photon and weak attenuated source, resulting in $\varepsilon = 0.92$ and $\varepsilon = 0.77$ respectively. Other factors to be considered include reflectivity and phase fluctuations, broadband filters, and the inefficiency of optical elements. However, as shown in the previous section, the influence of these factors is limited due to the pre- and post-selection processes; therefore, the deviation in the visibility and the experimental data points is negligible, such that the lowest confidence interval is $270$ for $\mu = 0.526$.

Our scheme differs from previously proposed schemes because it uses both the transmittance and the induced relative phase to identify the object's presence. Hence, it has high efficiency for detection and is robust against noise, a relative advantage compared to schemes that require multiple passes. However, it is essential to note that the efficiency metric $\eta$ used in previous research articles on interaction-free measurement differs from the probability of detection utilized in the current work.

\section{\label{sec:level6}Transmittance from Interference and Relationship to Weak Values}

The visibility can help us estimate the parameter $\mu$, which represents the transmittance of light through the intervening object. Using the quantum interrogation scheme proposed above, $\mu \geq 0$, $\varepsilon \geq 0$, Eq.~\eqref{eq:D-prob} and the bound of the cosine function, the visibility takes the form,
\begin{align} \label{eq:18}
    V(\mu, \varepsilon) &=  \frac{(1 + \mu + 2 \varepsilon \sqrt{\mu})/4 - (1 + \mu - 2 \varepsilon \sqrt{\mu})/4}{(1 + \mu + 2 \varepsilon \sqrt{\mu})/4 + (1 + \mu - 2 \varepsilon \sqrt{\mu})/4}  \nonumber \\
    &= 2 \frac{\varepsilon \sqrt{\mu}}{1 + \mu}.
\end{align}

\noindent The parameter can be estimated, therefore, if the purity of the input state is known. If a perfectly transmitting object is placed in the path, visibility simply reduces to the purity of the input state $(V \approx \varepsilon)$, while for a perfectly absorbing object, visibility drops to zero, which is expected as one of the arms in the interferometer is completely blocked.

To demonstrate this scenario, we have used a step-graded ND filter as the imperfect object. Unlike the continuous filter, the step filter ensures that for a certain translated position of the filter, the beam of photons with a sufficiently small diameter passes through a region of the object with constant transmission. Unfortunately, to determine this transmission constant $\mu$, we need to compensate for the phase $\delta$ and evaluate the $\varepsilon$ in the presence of the object in the arm.

Although adding a tilt to the BDP enables us to change the relative phase difference in the arms, after multiple attempts, we discovered that the phase introduced by the filter was too large to be compensated by the tilt without heavily compromising the photon count rate. To resolve this problem, we included another similar step-graded filter in the other arm of the interferometer, which transformed the nonunitary operator $\hat{A}$ to,

\begin{equation} \label{eq:Ap}
    \hat{A}^{'} = e^{i\delta}
    \begin{pmatrix}
        \sqrt{\mu_1} & 0 \\
        0 &  \sqrt{\mu_2}
    \end{pmatrix},
\end{equation}
where $\mu_1$ and $\mu_2$ are now the transmittance in each arm, and the phase can be ignored as it becomes global. Using the derivation from Sec.~\ref{sec:level3}, we find that detection probability after post selection takes the form,
\begin{align*} 
    D(\mu_1, \mu_2, \varepsilon, \phi) = \frac{1}{4} \bigg(\mu_1 + \mu_2 + 2 \varepsilon \sqrt{\mu_1 \mu_2} \cos(\phi)\bigg).\nonumber
\end{align*}
As before, we can evaluate the visibility using $\mu_{1} \geq 0$, $\mu_{2} \geq 0$, $\varepsilon \geq 0$ and the bound of the cosine function,

\begin{align} \label{eq:19}
    V(\mu_1, \mu_2, \varepsilon) = 2 \frac{\varepsilon \sqrt{\mu_1 \mu_2}}{\mu_1 + \mu_2}. 
\end{align}

\begin{figure}[]
    \centering
    \includegraphics[scale = 0.52]{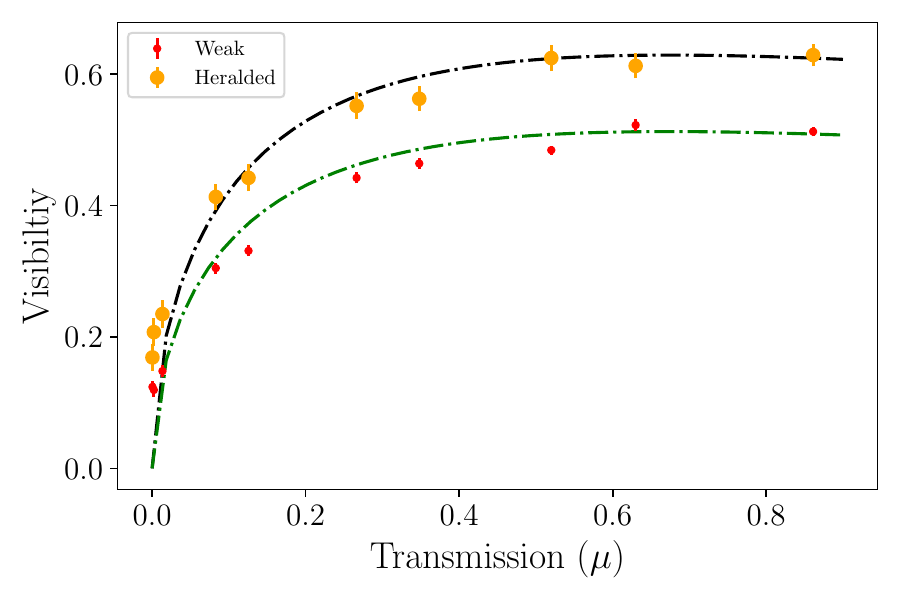}
    \caption{The visibility trend for the heralded single photon and coherent light shown by the yellow circles and red dots, respectively. The theoretical dot-dashed plots are also included, black for heralded single photons and green for the coherent light.}
    \label{fig: visibility_trend}
\end{figure}
\noindent We note that $\varepsilon$ can be independently determined using the transmittances of the objects in each arm of the interferometer. In particular, when the transmission constant in both arms is equal ($\mu_1 = \mu_2$), the expression for visibility in Eq.~\eqref{eq:19} reduces to $\varepsilon$.

When collecting experimental data for visibility trends against transmission, we kept transmission in one arm of the interferometer constant, $\mu_1 = 0.861$. By moving the graded filter in the other arm, we collected data for the visibility against transmission. The data shown in Fig.~\ref{fig: visibility_trend} for heralded single photons is given by the yellow diamonds, curve fit with the black dotted dashed line with $\varepsilon = 0.63$ in Eq.~\eqref{eq:19}. In comparison, the {coherent} light has lower visibility, plotted as red dots and curve fit with the green dotted dashed line with $\varepsilon = 0.51$. Therefore, by simply collecting data for visibility and $\varepsilon$ using the above-described steps, it is straightforward and somewhat reassuring to be able to determine the transmittance of the object, the accuracy of which is shown by the correlation of the fit with the experimentally collected data. Hence, with the addition of another object of known transmittance and phase, we can use the quantum interrogation setup discussed above to determine the transmission constant of the imperfect absorber using post-selection interferometry.

The experimental scheme we have followed may appear similar to what is conventionally done in a weak value measurement (WVM)~\cite{nirala2019measuring, rostiom22OptimalSettings}; however, despite structural similarities, there are notable departures. First, in a WVM, there is a weak interaction between a system of interest and a pointer degree of freedom, but in our arrangement, we squarely remain within the polarization system, and no interaction with a pointer is considered. In a typical WVM, which relies on pre-selecting a state $\ket{i}$ and after a feeble interaction with a pointer, projects the system onto a post-selected state $\ket{f}$, amplification in the pointer channel can be achieved if $\ket{f}$ is kept almost orthogonal to $\ket{i}$. In our case, the initial BDP creates an 
\begin{equation}
   \ket{i}=(\ket{H}+\ket{V})/\sqrt{2} \label{eq:i-state}
\end{equation}
(assuming $\varepsilon=1$) and the post-selector polarizer P is set at an angle $\theta$ not necessarily orthogonal, yielding
\begin{equation}
\ket{f}=\cos{\theta}\ket{H}+\sin{\theta}\ket{V}.\label{eq:f-state}
\end{equation}
Hence, there is no amplification. Nevertheless, our arrangement \emph{does} indeed employ pre- and post-selection, and here we show an interesting point that can be made in this context.

Previous work~\cite{nirala2019measuring} shows that through an interferometric arrangement, one can determine the expectation value of a non-Hermitian operator $\hat{F}$ which is decomposed in polar form as $\hat{F}=\hat{U}\hat{R}$ where $\hat{U}$ and $\hat{R}$ are, respectively, unitary and Hermitian operators and this decomposition is unique. The visibility of the fringes from the interferometer was shown to take the form,
\begin{equation} \label{eq:16}
        V = \frac{2 \bigl\vert \langle i | \hat{F} | i\rangle \bigr\vert}{1 + \langle i | \hat{R}^{2} | i\rangle},
\end{equation} 
where the expectation value in the pre-selected state $\langle i | \hat{F} | i\rangle$ can in general be complex and can be written in terms of the weak value of the Hermitian operator $\hat{R}$ \cite{pati2015measuring},
\begin{equation} \label{eq:17}
       \langle i | \hat{F} | i\rangle = \frac{\bra{f}\hat{R}\ket{i}}{\braket{f}{i}}\,  \braket{f}{i},
\end{equation} 
while noting that $\bra{f}\hat{R}\ket{i}/ \braket{f}{i}=\hat{R}_w$ is a weak value of the operator $\hat{R}$. An argument revolving around the theme of determining a weak value \emph{without} resorting to a typical WVM can be produced for our experiment as well.

Choosing $\hat{R}$ equal to the non-unitary but positive semidefinitive matrix $\hat{A}$ given in Eq.~\eqref{eq:A} and considering the initial and final states given in Eqs.~\eqref{eq:i-state} and \eqref{eq:f-state}, we show that 
\begin{equation}
\hat{A}_w(\theta,\delta,\mu)=\frac{1}{\sqrt{2}}(\cos{\theta}+\sin{\theta}\,e^{i\delta}\sqrt{\mu})
\end{equation}
and for $\theta=\pi/2$, we obtain,
\begin{equation}
\hat{A}_w(\theta=\pi/2,\phi+\delta,\mu)=\frac{1}{2}(1+\,e^{i(\phi+\delta)}\sqrt{\mu}),
\end{equation}
whose modulus square is identical to the probability of detection calculated in Eq.~\eqref{eq:D-prob},
\begin{align}
\bigl\lvert\hat{A}_w(\theta=\pi/2,\phi+\delta,\mu)\bigr\rvert^2 &= \frac{1}{4}(1+\mu+2\cos{(\phi+\delta)}\sqrt{\mu}), \nonumber \\
 &= D(\mu, \varepsilon=1, \phi+ \delta). 
\end{align} 
\noindent Hence, detection probabilities for our interferometric arrangement are equal to the (square of the) absolute values of the nonunitary positive semidefinite matrix defined for the partially absorbing object. Finally, the visibility will depend on the differences of the squares as given by,
\begin{equation}
V=\frac{\bigl\lvert \hat{A}_w\bigr\rvert^2_{\textrm{max}}-\bigl\lvert \hat{A}_w\bigr\rvert^2_{\textrm{min}}}{\bigl\lvert \hat{A}_w\bigr\rvert^2_{\textrm{max}}+\bigl\lvert \hat{A}_w\bigr\rvert^2_{\textrm{min}}}.
\end{equation}
\noindent This is equivalent to the visibility given in Eq.~\eqref{eq:v} and can therefore be used to derive Eq.~\eqref{eq:19}, which signifies that the transmittance of the object is equivalently determined using the weak value associated with the object.

\section{\label{sec:level7} Conclusions}
We have proposed a scheme for quantum interrogation that utilizes constructive interference to detect the presence of the object with high efficiency, even after a single pass. This scheme stands out from prior methods, as it uses the phase introduced by the object to detect its presence, irrespective of the transmittance. This means that, unlike prior methods, the proposed scheme can even detect the presence of an almost perfectly transmitting absorber, with limited losses.

We demonstrate that our scheme can work for both quantum and classical light by using a single photon and an unheralded coherent light source, respectively. We further analyze the dependence of the setup on the purity of the input state and its stability against noise from optical elements. We also show that with a small addition, we can adapt the setup to perform post-selection interferometry to determine the transmittance, allowing us to qualitatively extract information about the imperfect absorber. Finally, we demonstrate that our interrogation scheme returns the absolute weak values associated between pre- and post-selected states for a nonunitary operator, the imperfect absorber.

\begin{acknowledgments}
    We want to thank Wardah Mahmood and Muddasir Naeem for their insightful discussions.
\end{acknowledgments}

\bibliography{anwarJPPbib}

\end{document}